\documentclass[5p,sort&compress,twocolumn]{elsarticle}
\usepackage[utf8]{inputenc}
\usepackage[svgnames,hyperref]{xcolor}
\usepackage[final]{microtype}
\usepackage{xspace}
\usepackage{siunitx}
\usepackage{lineno}
\usepackage{hyperref}
\hypersetup{
    colorlinks=true,
}

\usepackage{amsmath}
\usepackage{amssymb}
\usepackage{miller}
\newcommand{\mean}[1]{\ensuremath{\left\langle {#1} \right\rangle}\xspace}
\newcommand{\fig}[1]{Fig.~\ref{#1}\xspace}
\newcommand{\equ}[1]{Eq.~\ref{#1}\xspace}
\newcommand{\dact}{\ensuremath{\Delta E_\mathrm{act}}\xspace}
\newcommand{\nicofeti}{Ni$_{36.67}$Co$_{30}$Fe$_{16.67}$Ti$_{16.67}$\xspace}
\usepackage{fancyhdr}
\fancypagestyle{pprintTitle}{\pagestyle{fancy}}
\begin{document}
\begin{frontmatter}
\title{Design using randomness: a new dimension for metallurgy}
\date{\today}
\author[add1]{W.G.~N\"ohring}
\author[add2]{W.A.~Curtin\corref{cor2}}
\ead{william.curtin@epfl.ch}
\address[add1]{Department of Microsystems Engineering, University of Freiburg, Georges-K\"ohler-Allee 103, 79110 Freiburg, Germany}
\address[add2]{Laboratory for Multiscale Mechanics Modeling, \'Ecole Polytechnique F\'ed\'erale de Lausanne, Lausanne, Switzerland}
\cortext[cor2]{Corresponding author}
\begin{abstract}
        High entropy alloys add a new dimension, atomic-scale randomness and the associated scale-dependent composition fluctuations, to the traditional metallurgical axes of time-temperature-composition-microstructure.  Alloy performance is controlled by the energies and motion of defects (dislocations, grain boundaries, vacancies, cracks, \dots).  Randomness at the atomic scale can introduce new length and energy scales that can control defect behavior, and hence control alloy properties.  The axis of atomic-scale randomness combined with the huge compositional space in multicomponent alloys thus enables, in tandem with still-valid traditional principles, a new broader alloy design strategy that may help achieve the multi-performance requirements of many engineering applications. 
\end{abstract}
\end{frontmatter}

\section{Motivation and perspective}

History has progressed through ages associated with materials, from the Stone Age to the Bronze Age to the Iron Age to the Silicon Age, but now we are in the Information Age.  A new metal age might, however, emerge due to a revolution in metallurgy.  In spite of incredible advances in information and artificial intelligence, society exists in a physical world, and that physical world runs on energy and is threatened by energy issues.  To reduce energy use and emissions requires efficient energy use and generation, and these in turn require higher-performance -- more efficient, more durable, stronger, lighter, safer -- materials for catalysts, batteries, solar energy conversion, hydrogen storage, but also for structures used mainly for transportation.  While advanced composites and novel nanomaterials can fill some key needs, many broad requirements for structural materials are well-met by metal alloys.  Advanced metals, such as new steels and engineered Aluminum, Magnesium, and Superalloys, are making accelerated progress toward essential goals, assisted now by exceptional new experimental tools such as Atom Probe Tomography and immensely greater computing power that both yield quantitative information at the atomic scale.  The emergence of a new metals age may occur, however, due to the confluence of these advances with the discovery of a new class of ``High Entropy Alloys'' (also ``Compositionally Complex Alloys'' or ``Multiple Principle Element Alloys'') that burst into full view only a decade ago.  

Multicomponent HEAs to date have families that overlap with traditional steels (Fe-Co-Ni-Mn-Cr), refractory metals (Cr-Mo-Nb-Ta-V-W), lightweight metals (Al-Ti-Sc-Mg-Li), precious metals (Rh-Ir-Pt-Pd-Au-Ag-Ni-Cu), and others \cite{miracle_critical_2017,gorsse_mapping_2017,li_strong_2017}.  Initial studies have mainly been on equicomposition alloys, but there is no restriction to this special case.  The composition space is immense: a 5-component alloy has over 650,000 compositions at composition increments of ~1.6\%.  Which among these 650,000 possibilities is the strongest?  The lightest?  Retains the highest strength at high temperatures?  Is most resistant to oxidation, creep, grain growth, fracture or Hydrogen embrittlement?  Are there compositions that would satisfy many of these needs?  What would be the properties upon dilute addition of a 6th alloying element? Materials scientists bring vast intuition into selecting among possibilities to optimize various properties based on traditional concepts.  But, even with high-throughput experiments, the full range of compositions cannot be evaluated.  If a new metal age is to emerge, materials scientists must gain quantitative control over the vast space of possible HEAs.

The key feature of HEAs is that, while crystalline, they have the individual atomic sites occupied (essentially) randomly by each atom type.  That is, HEAs are compositionally disordered at the very scale of the atomic spacing.   Traditional materials science is largely based on understanding the behaviour of crystalline defects -- dislocations, solutes, precipitates, grain boundaries, twins, and polycrystalline texture -- in an underlying ``matrix'' \cite{argon_strengthening_2008,kocks_thermodynamics_1975}.  The defects often play a very positive role for many properties, particular enabling the plastic flow and high fracture toughness that are the dominant mechanical advantages of metals over other structural materials.  The controlled introduction of defects improves the metal, ultimately determining the combination of properties (strength, hardening, ductility, creep resistance, fracture toughness, fatigue resistance, all as a function of temperature) that allow metals to be tailored for different applications.  Paraphrasing F.\ C.\ Franck, ``crystals are like people; it is the defects that make them interesting''.  More importantly, it is the defects that make them useful.  In HEAs, there is no clearly-defined matrix: \emph{all atoms are ``defects''} in a broad sense.  In HEAs, the standard defects (dislocations, grain boundaries, interfaces, cracks) are actually defects existing within a material that has atomic scale defects everywhere (the atoms themselves).  This presents an entirely new conceptual problem: \emph{how do we understand the structure and behaviour of the traditional defects in a material that is random at the atomic scale? but it also presents an opportunity: can we use this new feature of atomic scale randomness to achieve improved performance?}  

Materials science textbooks and literature typically consider ``average'' properties: an alloy is mainly another metal material, or metal/intermetallic mixture, with its particular crystal structure, lattice constants, elastic constants, stacking fault energies, etc.\  Design of new materials is then guided by controlling average properties of accessible phases.  There are exceptions but the general neglect of randomness that may miss the controlling energetics of many phenomena.  Thus, while it is experimentally established that the basic mechanisms and defects operating in HEAs are essentially those of standard alloys \cite{george_high_2020}, the behaviors of these defects can be modified in essential respects due to the high degree of atomic randomness. \emph{The challenge is then to understand how/if the high complexity of the HEAs changes the operation of these mechanisms.}

How can the spatial randomness in HEAs can be exploited to enhance a range of material properties?  Spatial randomness occurs over all length scales. Consider an $n$-component random alloy at composition {$c_i$} ($i=1,\dots{}n$).  In a volume of material containing $N$ atomic sites, there are on average $c_iN$ type-$i$ atoms at concentration $c_i$.  But, there are also fluctuations scaling as $\propto\sqrt{c_iN}$ in the actual number of type-$i$ atoms around the average value (Fig. \fig{fig:schematic}a).  A defect in this region of $N$ atoms interacts with the actual solutes, not the average, and so the energy of the defect is changed, relative to the average, by its interactions with these $\sqrt{c_iN}$ fluctuations.  Therefore, random solute fluctuations change the defect energy at all scales (since $N$ is arbitrary) -- the defect energy becomes intrinsically scale-dependent.  The defect responds (moves or changes structure) to these scale-dependent energy fluctuations so as to lower the total energy of the system.  There is some energy cost to create the structural changes, and hence an energetic balance is struck.  The interaction of the solute fluctuations with the defects then creates new composition-dependent length and energy scales in the random alloy. The behavior of the defect is then controlled by these solute-induced material length scales and their associated energy scales.  Specifically, the motion and behaviour of the defect is often, but not always, inhibited: the defect must move from the lower-energy structure through nearby higher-energy structures (unfavourable fluctuations), increasing the barriers and stresses for the underlying defect motion.  Solute fluctuations change the energetics of different defect processes to different degrees, depending on the nature of the defect(s) involved.  In other words, the fluctuations affect point defects, dislocation line defects, crack-front line defects, and planar grain boundary defects, but affect each defect in its own way and at its own scale.

\begin{figure}[htb]
    \centering
    \includegraphics{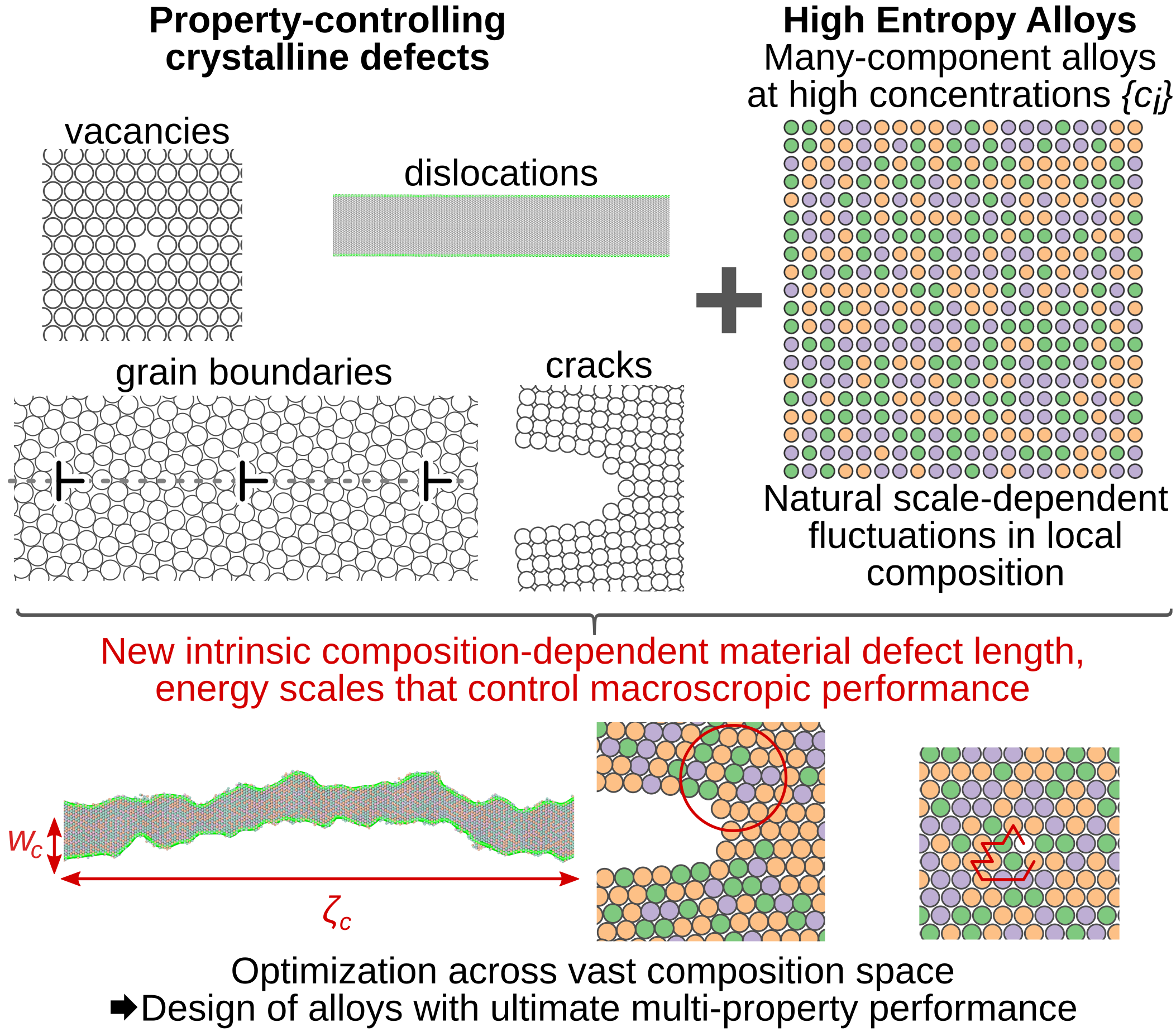}
    \caption{The behavior of standard crystalline defects in metals with the intrinsic random arrangement of atoms in an HEA leads to new length and energy scales in the material which, in turn, modifies the defect structure, energy, and motion, thus influencing macroscopic properties.  Understanding the connections between randomness and defects can enable the selection of alloy compositions that could optimize multiple alloy properties.  
    \label{fig:01}
    }
\end{figure}

\section{Conceptual framework}

\begin{figure}[htb!]
    \centering
    \includegraphics[width=\columnwidth]{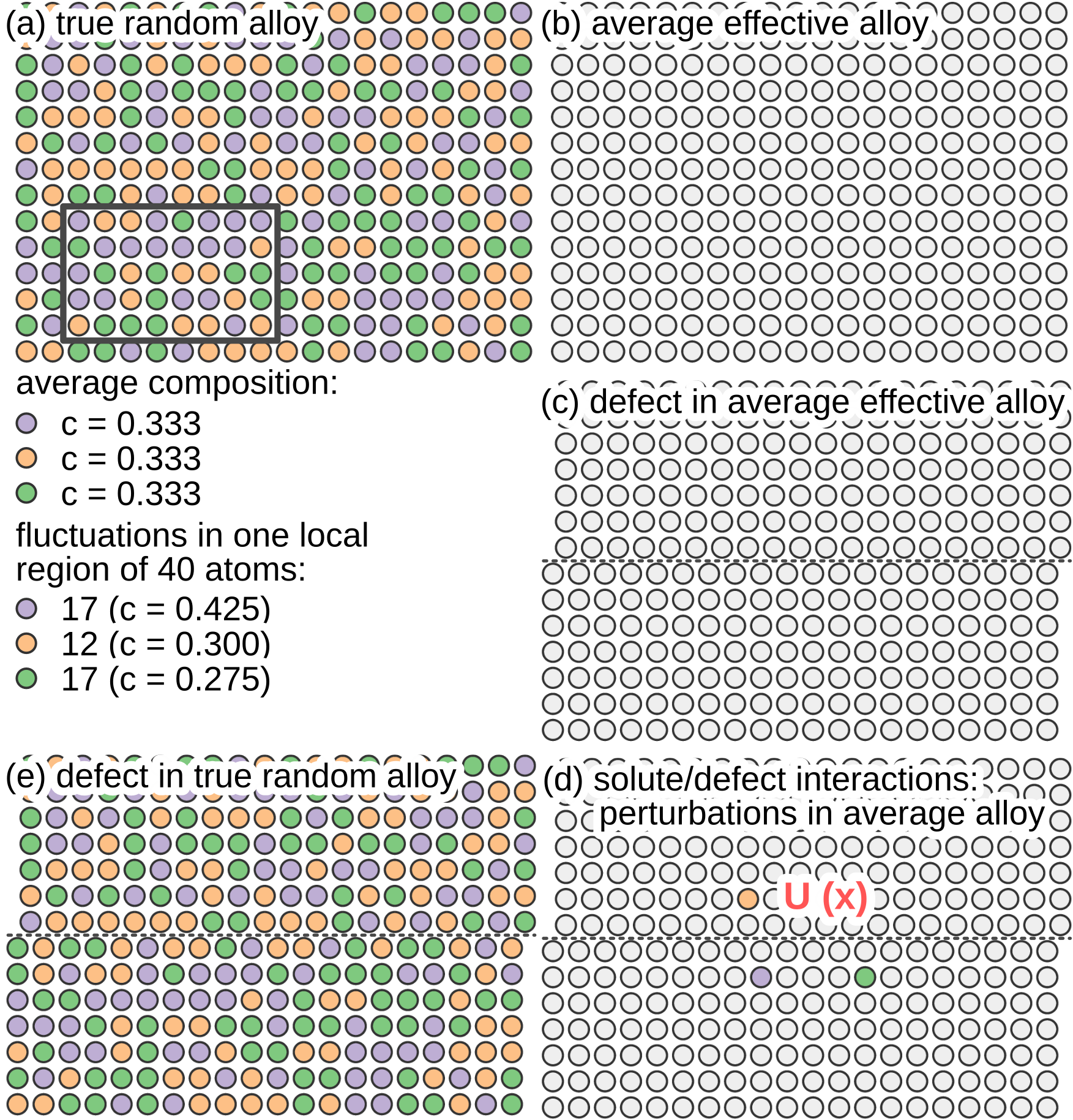}
    \caption{
    \label{fig:schematic}
    Systematic approximation to solute-defect binding energy fluctuations using the average- or A-atom approach. 
    The true random alloy (a) is replaced by an equivalent configuration of A-atoms (b) to model 
    the behavior of the average matrix. (c) A defect (here, a stacking fault) is inserted into the 
    A-atom configuration. By replacing individual A-atom with true atoms (d), the solute-defect
    interaction energy can be computed for every site and every solute type. These values can be used
    to estimate energy fluctuations due to local solute fluctuations in the true random alloy containing the defect (e).
    }
\end{figure}

To understand how fluctuations control properties requires a conceptual framework as follows.  We must imagine that the actual random alloy has an average homogeneous counterpart.  Instead of thinking about each atom type X, Y, Z, \dots{} as distinct, we envision an underlying ``average atom'' A for the alloy (Figure \fig{fig:schematic}b).  The alloy composed only of these A atoms then has exactly the same average properties as the random XYZ\dots{} alloy -- the same lattice constant, the same elastic constant, the same stacking fault energies, the same surface energies, the same grain boundary energies, the same dislocation core energies -- i.e.\ all properties that are averages over a large volume of alloy material are preserved.  The A atom represents the collective average of the interacting X, Y, Z,\dots{} atoms at the alloy composition. Furthermore, each actual atom X, Y, Z, \dots{}. can then be viewed as a solute added into the A material, replacing an A atom.  The individual X, Y, Z,\dots{} atoms introduced into the A material are then \emph{perturbations} of the material away from the average; all of the average interactions among X, Y, Z,\dots{} atoms are \emph{already incorporated} into the A atoms. The A atom can be explicitly constructed for atomic systems described by interatomic potentials, and the idea underpins the Coherent Potential Approximation in first-principles methods, but it is the general concept that is important. 

Now consider a crystalline defect in the true random X-Y-Z\dots{} alloy; for simple illustration, we consider a stacking fault as shown in \fig{fig:schematic}c.  The role of randomness on local energetics of the stacking fault is then assessed as follows.  First, the stacking fault is created in the A-atom material, and has an energy $E^{SF}_A$.  Then the individual actual atoms (X, Y, Z\dots{}) can be introduced into the distinct atomic sites around the fault (\fig{fig:schematic}d).  An atom of type X introduced into site $j$ of the stacking fault in the A material has some interaction energy $U_X^{SF}(j)$ (i.e.\ the energy change of atom X when moved from a position far away from the fault to the site $j$ near the fault).  The concentration-weighted sum over all atom types $\sum_{X, Y\dots{}}  c_X U_X^{SF}(j) = 0$; there is zero average effect because the A material already represents the average.  The real random material is then the A material with a specific atom (X, Y, Z,\dots{}) introduced into each and every site in the material (in and away from the grain boundary); this establishes an approximation to the true random alloy with the defect (\fig{fig:schematic}e).  Assuming that the individual solutes X, Y, Z,\dots{} do not interact with one another (beyond the overall average interactions that establish the A material), the total energy of the system is 
\begin{align}
E_{TOT} = E^{SF}_A + \sum_j^N \sum_X^{n} s_{jX} U_X^{SF}(j)
\end{align}
where the occupation variable $s_{jX}=1$ if an X atom resides at site $j$ and $=0$ otherwise.  The average energy is 
\begin{align}
\begin{aligned}
\mean{E_{TOT}} &= E^{SF}_A + \sum_j^N \sum_X^{n} \mean{s_{jX}} U_X^{SF}(j) \\
               &= E^{SF}_A + \sum_j^N \sum_X^{n} c_X U_X^{SF}(j) = E^{SF}_A \\
\end{aligned}
\end{align}
as stated above.  However, in any specific finite region of the actual random material, there are statistical fluctuations in the energy.  The variance (square of the standard deviation) of those energies can be computed as 
\begin{align}
    \sigma^2 = \mean{\sum_{i,j}^N \sum_{X,Y}^{n} s_{iX}s_{jY} U_X^{SF}(j)U_Y^{SF}(i)}  \label{equ:sf}
\end{align}

Locally, the energy of the SF is varying around the average value.  In some regions, the specific arrangement of solutes lowers the SF energy.  In other regions, however, the specific arrangement of solutes increases the SF energy.  The standard deviation in energies depends on the area of SF considered, which involves some $N$ atoms (area $x$ thickness over which solute/SF interactions are non-negligible), and so scales with $\sqrt{N}$.  Larger areas, while having smaller fractional deviations from the average, have larger energy fluctuations in absolute value.  Properties that depend on the stacking fault energy now become dependent on the fluctuations in energy of the stacking fault over some scale associated with the defect of interest.  If the defect moves, e.g.\ the stacking fault is formed on another plane due to cross-slip (see example below), then there is a standard deviation for the change in energy,
\begin{align}
    \sigma = \sqrt{\sum_{j}^N \sum_{X}^{n} c_{X}(1-c_{X}) \left(\Delta U_X^{SF}(j)\right)^2} \label{equ:stddev}
\end{align}
where $\Delta U_X^{SF}(j)$ is the change in energy of the X atom at atomic site $j$ due to motion of the stacking fault, which changes position of site $j$ \emph{relative} to the SF.

\section{Illustrative example: cross-slip in an FCC HEA}

\begin{figure}[htb]
    \centering
    \includegraphics{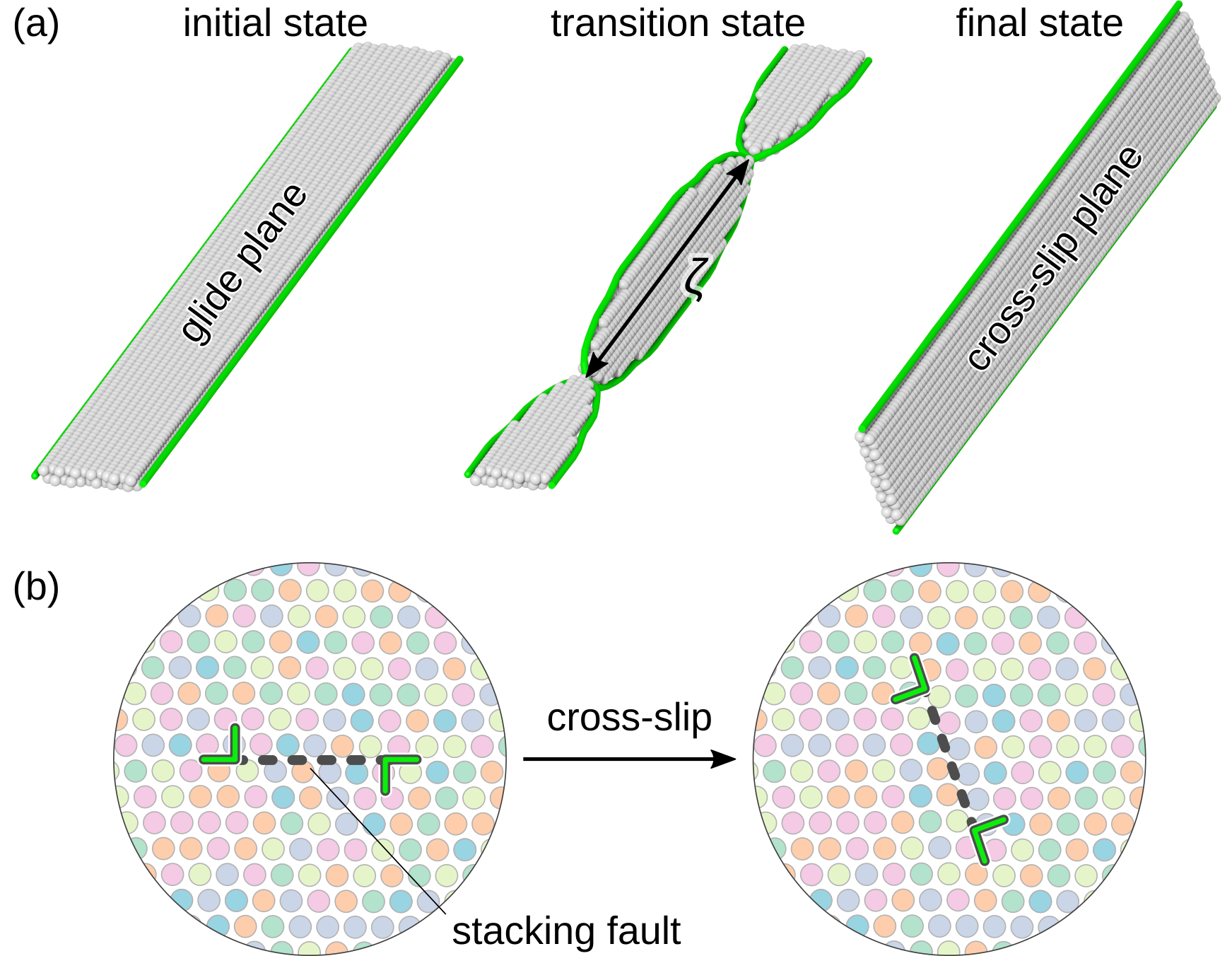}
    \caption{(a) Transition path for fcc dislocation cross-slip by the Friedel-Escaig mechanism; green: Shockley partial dislocations; white: atoms in the stacking fault of the dislocation; the transition state has a characteristic length $\zeta$; (b) In a fixed atomic environment (2d schematic), the distribution of atoms relative to the dissociated dislocation changes upon cross-slip, and there is an associated energy change for this particular arrangement of atoms.  This energy change affects the cross-slip barrier in this region of the material.
    \label{fig:crossslipschematics}
    }
\end{figure}

The SF case above can have a specific impact on various alloy properties.  In particular, here we consider cross-slip of a screw dislocation in an fcc alloy.  A lattice screw dislocation in an fcc crystal dissociates into two partials separated by SF of width $d$.  To cross-slip onto a different plane, the two partials must re-combine (constrict) into a compact core and then re-dissociate onto the new plane via the well-known Friedel-Escaig.  The process is thermally-activated, with a transition state corresponding to a cross-slip nucleus having two constrictions and a region already cross-slipped onto the cross-slip plane, see  \fig{fig:crossslipschematics}(a).  The average energy barrier for this process is that for the corresponding A-atom average alloy, and there is a corresponding transition state configuration with a characteristic length $\zeta$.  The average energy barrier scales as $\mu b / \gamma_{sf}$, showing that elements and alloys with low average stacking fault energy have very high average cross-slip barriers and hence very low average cross-slip rates at room temperature.  Applied  stresses can reduce the barrier and thus enable cross-slip, but high stresses are needed in materials with low SF energy.  

In the random alloy, the random composition fluctuations introduce an additional contribution to the energy barrier because there are local differences in the energies of the original stacking fault, the transition state, and the final state over the length $\zeta$.  This local energy is determined by the local atomistic configuration in some tubular-shaped domain of size $\propto \zeta w^2$, where $w>d$ is the relevant radius.   In this case, the solute fluctuations do not change the length scale ($\zeta$) but do lead to energy fluctuations.  One contribution to these energy differences is due to fluctuations in the solute/SF interaction energy, as characterized by the standard deviation of \equ{equ:stddev} above at the scale $d \zeta$.  This contribution is embedded within the total fluctuations due to the solute/dislocation interaction energies, see \fig{fig:crossslipschematics}(b) \cite{nohring_dislocation_2017,nohring_cross-slip_2018}.  In HEAs, all atoms are solutes relative to the average A material, and hence the fluctuations could be large, and the consequences for cross-slip far more significant than in dilute alloys.

\begin{figure}[htb]
    \centering
    \includegraphics[width=\columnwidth]{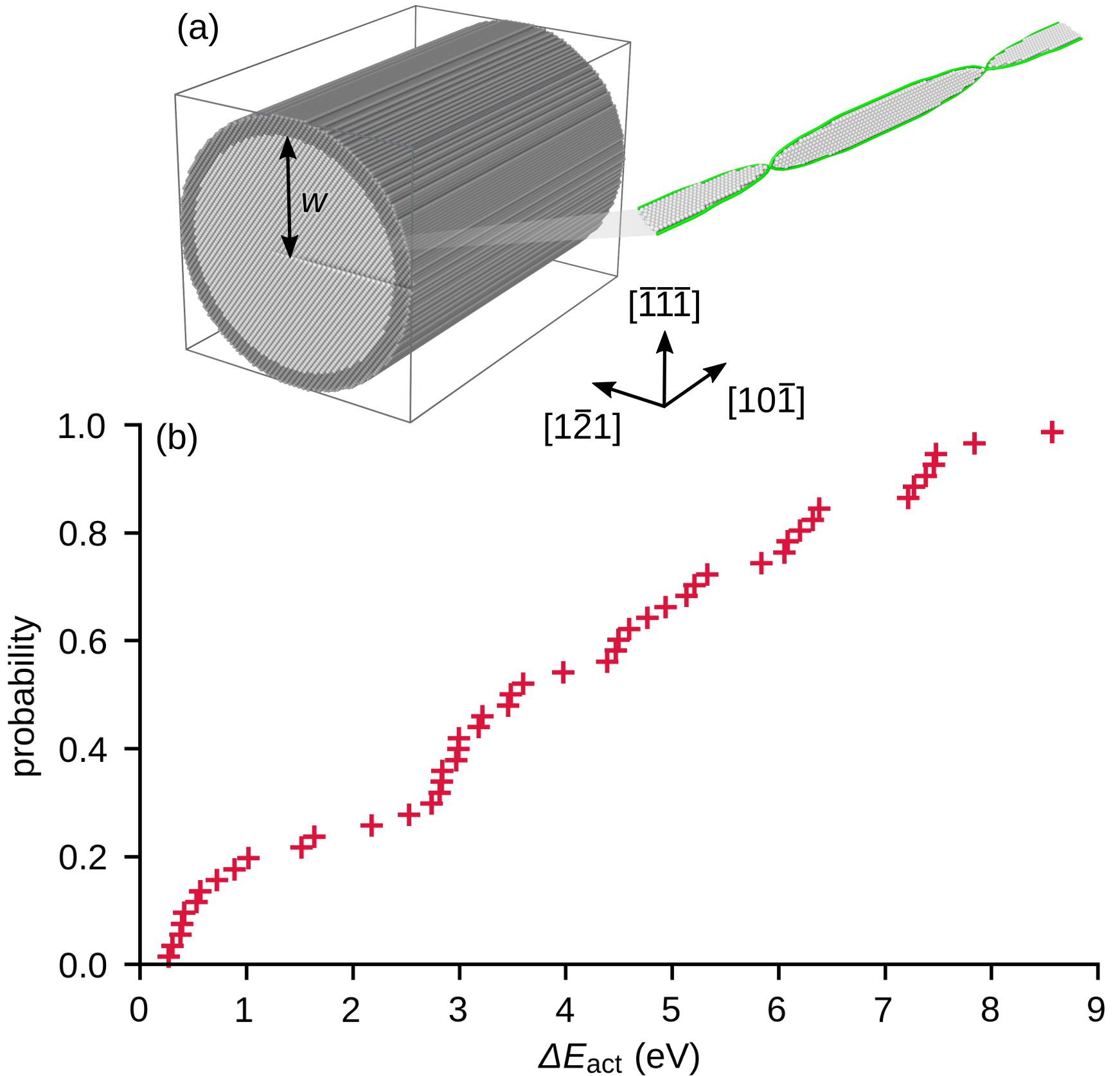}
    \caption{(a) Configuration used for calculating activation energies \dact; the 
    cylinder has length $\zeta=130b$ and inner radius $w=15\sqrt{3}a$; the dislocation
    is at the center, parallel to the cylinder axis; atoms in
    the shell (dark, width equal to two times the cutoff radius of the potential) 
    are fixed  (b) Cumulative probability of \dact across 49 different random solute distributions; approximate probability $p=(i-0.3)/(n+0.4)$ \cite{benard_het_1953}, with sample size $n=49$ and $i=1\dots{}n$.
    \label{fig:crossslip}
    }
\end{figure}

For example, Rao et al.\ simulated screw dislocation glide in a model \nicofeti HEA. They observed an essentially-spontaneous cross-slip event at \SI{300}{\kelvin} \cite{rao_atomistic_2017} even though the A-atom stacking fault energy is very low ($\gamma_{sf}=\SI{14.8}{\milli\joule\per\meter\squared})$ and the cross-slip barrier very high $\SIrange[range-phrase=\text{--},range-units=single]{4.6}{4.9}{\electronvolt}$. We have computed the distribution of cross-slip energy barriers in this same random alloy using atomistic minimum 
energy path calculations (the modified String method; see details in \cite{nohring_dislocation_2017,nohring_cross-slip_2018}).  Briefly, we consider a straight dislocation in a tubular domain of length $2\zeta=130b$ and radius $15\sqrt{3}a$ that is sufficient to capture the transition state and the overall energetics with good accuracy, see \fig{fig:crossslip}(a).  This simulation cell size is slightly too small for highly accurate results, but is necessary to reduce computational cost and our main conclusions are not affected.  The initial state corresponds to a dislocation dissociated on the \hkl{-1-1-1} glide plane and the final state corresponds to a dislocation dissociated on the \hkl{-11-1} plane with the same center position as the initial state (see also \fig{fig:crossslipschematics}(b)).  The minimum energy path between these configurations was calculated for many different atomistic realizations of the true random alloy.  Even during simulation sample preparation, we observed spontaneous cross-slip in 6 cases, i.e.\ the dislocation spread on both initial and final planes during the initial energy minimization phase. We obtained
 49 cases in which the initial and final configurations had the dislocation spread completely on
 the glide and cross-slip plane, respectively.   \fig{fig:crossslip}(b) shows the cumulative probability of the energy barriers across the 49 cases with a non-zero barrier. The mean energy barrier \SI{3.84}{\electronvolt} is very large as expected  (although not quite equal to the average-atom value).  More importantly, there are very large 
 fluctuations in the energy barrier, including very low and very high barriers.  These fluctuations are due to the \emph{specific} random atomic arrangements around the dislocation in each individual random realization, which leads to large fluctuations in both solute-dislocation and solute-solute interaction energies between the initial and final states.  In some cases, $\Delta E_\mathrm{act}$ is very close to zero, and $\sim$ 10\% of the cases have a barrier lower than 0.5 eV (comparable to Al that is considered to cross-slip very readily).  A dislocation encountering these random low-barrier environments will have a very high cross-slip rate --- almost spontaneous --- in spite of the fact that the barriers elsewhere are much higher.  Once cross-slip is initiated, it can spread along the entire length, particularly if assisted by a (small) Schmidt stress on the cross-slip plane \cite{nohring_cross-slip_2018}.   These findings are
 consistent with the observation of spontaneously cross-slip configurations in the simulations of Rao et al.  In spite of the immense and essentially insurmountable average cross-slip barrier in this alloy, cross-slip can occur with nearly zero energy barrier due to favorable local fluctuations of the solutes on the cross-slip plane relative to the initial glide plane.  Experimentally, possible evidence for 
 fluctuation-assisted cross-slip is found in Ref. [Ding et al.\ \cite{ding_real-time_2019}], where the mechanical behavior in CoCrNi, CoCrFeMnNi, and  CoCrFeNiPd HEAs at cryogenic temperatures was studied via in-situ Transition Electron Microscopy. The authors observe 
 extensive cross-slip in all three alloys, with in the average stacking fault energy having a minor role. Such insensitivity is expected if the rate of cross-slip is controlled by local fluctuations in the solute concentration.

\section{Randomness and defects: future directions}

We now return to the general philosophy outlined in the first section of the paper, and discuss the implications of randomness on the various defects known to control properties in metals.  Several new directions and ideas are put forth.

\subsection{Line defect: dislocations and plasticity}

We have already made significant progress in modeling the initial yield stress in fcc and bcc HEAs \cite{varvenne_theory_2016,varvenne_solute_2017,yin_first-principles-based_2019,maresca_theory_2020,maresca_mechanistic_2020} using the general framework above.  Dislocations in a random alloy spontaneously become wavy at some characteristic scale (wavelength $\sim 4 \zeta_c$, amplitude $w_c/2$) so as to minimize the total energy.  The dislocation is then residing in low-energy regions, and motion (plastic flow) requires the dislocation to move over the barriers created by the adjacent high-energy regions, leading to a temperature- and rate-dependent yield stress.  The theories provide very good predictions in comparison to recent experiments (see above references and others by the same authors).

Twinning in fcc HEAs has been found to operate at stresses much higher than expected based on correlations of twinning stress and stable stacking fault energy.  The twinning is speculated to play a major role in hardening of local high-stress regions to prevent the local onset of failure, enabling high failure strains \cite{laplanche_reasons_2017}.  Control of the twinning stress is thus valuable.  We postulate that, unlike in the elements and dilute alloys, there is high ``solute strengthening'' of the twin dislocations in HEAs that will control the operative strength.  That is, twin nucleation may be relatively easy but growth (motion of twin dislocations along the initial twin nucleus) will determine the operative strength.  The analysis would follow the lines of analysis for lattice dislocations, with modifications associated with the twin dislocation structure and its constraint to the twin boundary. 

Theory to date has neglected the effects of short-range order and explicit solute-solute interactions on dislocation motion and flow stress.  In the random alloy, solute-solute interactions present another source of fluctuations associated with the creation and destruction of solute-solute pairs (at any distances) across the glide plane.  We have now computed these fluctuations for fcc and bcc alloys, and can predict the influence of solute-solute fluctuations on the yield stress \cite{nag_notitle_nodate}.  Results in a bcc HEA indicate that solute-solute interactions on the order of \SI{0.1}{\electronvolt}, which is fairly large for thermodynamic and phase stability issues, has quite a small effect on yield strengths.  However, recent data on the AuNiPdPt alloy shows a strength much higher than predicted by the current theories based only on solute-dislocation interactions \cite{thiel_breakdown_2020}, which might be attributable to large interaction energies of Au with the other alloying elements.

In the bcc alloys, an initially straight screw dislocation will also spontaneously form a kinked structure so as to lower its total energy (gaining potential energy due to interactions with favourable local solute environments at the cost of the kink formation energy).  Thus, the randomness introduces a new length scale associated with the density of kinks in the random material.  The same process leads also to the formation of cross-kinks --  when one screw dislocation kinks/glides on different glide planes at different positions along its length.  This sets a second related material length scale due to the randomness.  Strengthening then emerges from a combination of kink glide through the random solute field and the need to break the ``cross-kinks'', which are strong obstacles with high energy barriers.  In non-dilute alloys and HEAs, cross-kinks can be far more frequent, and thus may account for the observed retention of strength at high temperatures.  We have further recently shown that edge dislocations, usually neglected in bcc alloys, may have strengths comparable to, or higher than, those of screw dislocations, especially at high temperatures.  In this case, the high randomness can lead to a fundamentally different mechanism controlling yield strength in complex bcc alloys, representing a clear departure from the traditional understanding.

\subsection{Point defects: vacancies and H diffusion}

At high temperatures, creep deformation can govern mechanical performance because vacancy concentrations and diffusion rates becomes appreciable. The interaction of point defects with the solute fluctuations in HEAs may have a significant effect on vacancy diffusion.  The activation enthalpy for diffusion of a species is a sum of the vacancy formation enthalpy and a solute-specific migration enthalpy, $\Delta{}H_i =H_f + \Delta{}H_{m,i}$.  While the vacancy formation energy is an average quantity across the entire system, the local migration enthalpy is a function of local composition, and the formation enthalpy at any local site varies around the average, contributing to the local barrier for diffusion.   Larger-scale composition fluctuations can generate scale-dependent internal stresses that couple to the vacancy misfit strain tensor.  These variations give rise to additional energetic barriers for diffusion over various scales, possibly related to so-called “super basins” \cite{trinkle_notitle_nodate}.  Recent investigations using first-principles MD in small systems have further indicated that there are some ``percolation-like'' aspects to diffusion in HEAs \cite{zhao_preferential_2017}.  That is, there is one most favourable solute type for vacancy migration, and the vacancy migration can thus occur mainly by diffusion along connected chains of the favourable solute type.  If the favourable solute concentration is below the percolation threshold for the lattice structure, then the vacancy is forced to exchange with less-favorable solutes, lowering the diffusion rate.  This concept has not been fully developed and may require further modifications as larger length scales are considered.  Another avenue for envisioning vacancy diffusion in complex alloys comes from long-standing models of electron ``hopping'' transport in disordered materials such as amorphous Si.  Diffusion in systems with a broad distribution of traps for the diffusing species leads to so-called anomalous diffusion.  This has been understood within the framework of the Continuous Time Random Walk (CTRW) model \cite{scher_stochastic_1973,scher_anomalous_1975}.  Concepts from these models may thus be applicable to HEAs.

Hydrogen, as an interstitial, will also have a complex energy landscape in a complex alloy.  A distribution of local H absorption energies in the lattice leads to trapping of H in the lowest-energy sites throughout the alloy.  This can inhibit the H atoms from aggregating at dislocations or crack tips, for example, and thus inhibit Hydrogen embrittlement.  Recent experiments show that CoCrFeMnNi and CoCrFeNi are more resistant to H embrittlement than Ni and stainless steels, even while absorbing more H under the same charging conditions.  This is consistent with the conceptual picture of local trapping of H throughout the complex alloy preventing/slowing possible mechanisms thought to drive embrittlement.

\subsection{Planar defects: grain boundaries}

At high temperatures, whether during application or fabrication, polycrystalline alloys can coarsen by grain growth.  In alloys, the segregation of solute elements to the GB can slow grain growth at high temperatures, and there is a well-established solute-drag phenomenon.  We envision here a very different mechanism in HEAs that does not require solute diffusion/segregation, and therefore may suppress grain growth without undesirable segregation.  As with other defects, the spatial fluctuations in solute concentration in the random alloy should create spatial variations in local grain boundary (GB) energy.  The GB plane itself can then fluctuate locally to find low energy undulating configurations.  There is, however, an additional energy cost to create steps/disconnections.  The competition between the lowering of the potential energy and the increasing energy due to disconnections will establish characteristic length and energy scales for the lowest-energy ``wavy'' GB structure.  GB motion then requires additional thermal activation out of the low-energy state and through the nearby high-energy configurations.  Since grain coarsening generally has a small driving force, this pinning of the GB could be expected to strongly suppress grain growth.    This concept is a direct analogy to the strengthening of dislocations, but applied the grain boundary.

\subsection{Multi-dimensional defects: cracks and intrinsic ductility}

Fracture in metals is a complex process, with most metals failing by mesoscale ductile fracture modes that are not intrinsically atomistic.  However, some processes such as fatigue crack growth and H embrittlement involve very sharp cracks, and so the behavior at sharp cracks can be important.  In addition, materials are deemed intrinsically brittle or intrinsically ductile based on whether an atomically-sharp crack will grow by cleavage (brittle) or will emit dislocations and blunt (ductile).  In general, for a given geometry, there is an applied stress intensity for cleavage ($K_{Ic}$) (the Griffith criterion) and for dislocation emission ($K_{Ie}$).  Intrinsically ductile materials have $K_{Ie} < K_{Ic}$.  The lower the value of $K_{Ie}$, the more difficult it is for sharp cracks to grow.  In HEAs, the crack front encounters solute fluctuations, both along the line and along the putative emission slip plane and fracture plane.  Brittle fracture requires global thermodynamics to be satisfied, and thus macroscopic crack growth cannot decrease below the Griffith value corresponding to the average material surface energy for the chosen cleavage plane.  In contrast, the dislocation emission instability can be nucleated at local regions of low unstable stacking fault energy. Thus, HEAs hold the potential for significant reductions in $K_{Ie}$ relative to $K_{Ic}$.  The scale of these reductions is related to the energy fluctuations that occur over the scale of the nucleating loop (in the 3D process).  It may become possible that a material that is brittle on average ($K_{Ic} < K_{Ie\mathrm{(average)}}$) maybe ductile due to the local compositional fluctuations.  A competing negative effect in HEAs is, however, that the alloy resists dislocation motion (high flow stress).  Thus, it is difficult for nucleated dislocations to move far from the crack tip region.  The back-stresses due to nucleated dislocations inhibit further nucleation and can also increase opening stresses that drive cleavage failure.  The behaviour of cracks in HEAs is thus opens several new complications relative to simpler materials.  

\section{Closing remarks}

The objective of this viewpoint article has been to highlight the fact that the inherent high randomness at the atomic scale ---  the dominant feature of high entropy alloys --- naturally introduces a host of interesting new considerations beyond the scope of traditional metallurgy.  Specifically, the interaction of traditional defects with the atomic-scale randomness can give rise to new length and energy scales, and these length and energy scales can quantitatively and/or qualitatively change the behaviour (motion, evolution) of the defects.  Since it is the defect behaviors that control properties, the randomness then can have a fundamental impact on properties beyond those represented by the average properties of the alloy.  We have shown one new application of this concept in our example of cross-slip, where an immense average cross-slip barrier can be reduced to near-zero in some local regions, sufficient to nucleate cross-slip, consistent with recent simulations and observations of significant cross-slip in several HEAs having low stacking fault energies.  The overall influence of randomness has then been discussed in the context of dislocations, vacancies and Hydrogen, grain boundaries, and cracks -- the dominant crystalline defects that control macroscopic mechanical properties -- and new ideas for pursuing future research about the role of atomic scale randomness on these defects have been offered.  With such understanding, the dominant feature of randomness in HEAs may be harnessed to design new complex alloys with multi-property performance, providing a significant advance in metallurgical science.

\section{Acknowledgments}
This work was supported by the Swiss National Science Foundation project ``Harnessing atomic-scale randomness: design and optimization of mechanical performance in High Entropy Alloys'', project ``200021\_18198/1''.
WGN acknowledges support from the  European Research Council (ERC-StG-757343).

\appendix
\bibliographystyle{elsarticle-num}
\bibliography{2020_scripta_article_design_with_randomness_references}

\end{document}